\documentclass[letterpaper]{article} 
\usepackage{aaai24}  
\usepackage{times}  
\usepackage{helvet}  
\usepackage{courier}  
\usepackage[hyphens]{url}  
\usepackage{graphicx} 
\usepackage{booktabs}
\usepackage{natbib}  
\usepackage{caption} 
\usepackage{algorithm}
\usepackage{algorithm}
\usepackage{algorithm,algpseudocode}
\usepackage{amsmath}
\usepackage{amsfonts}

\usepackage{newfloat}
\usepackage{listings}
\DeclareCaptionStyle{ruled}{labelfont=normalfont,labelsep=colon,strut=off} 
\floatstyle{ruled}
\newfloat{listing}{tb}{lst}{}
\floatname{listing}{Listing}
\title{\emph{SentinelLMs}: Encrypted Input Adaptation and Fine-tuning of Language Models for Private and Secure Inference}
\author {
    Abhijit Mishra, Mingda Li, Soham Deo
}
\affiliations {
    School of Information \\
    University of Texas at Austin \\
    {abhijitmishra, mingdali, soham.deo}@utexas.edu
}

\usepackage{bibentry}

\begin{document}

\maketitle

\begin{abstract}
This paper addresses the privacy and security concerns associated with deep neural language models, which serve as crucial components in various modern AI-based applications. These models are often used after being pre-trained and fine-tuned for specific tasks, with deployment on servers accessed through the internet. However, this introduces two fundamental risks: (a) the transmission of user inputs to the server via the network gives rise to interception vulnerabilities, and (b) privacy concerns emerge as organizations that deploy such models store user data with restricted context. To address this, we propose a novel method to adapt and fine-tune transformer-based language models on passkey-encrypted user-specific text. The original pre-trained language model first undergoes a quick adaptation (without any further pre-training) with a series of irreversible transformations applied to the tokenizer and token embeddings. This enables the model to perform inference on encrypted inputs while preventing reverse engineering of text from model parameters and intermediate outputs. After adaptation, models are fine-tuned on encrypted versions of existing training datasets. Experimental evaluation employing adapted versions of renowned models (e.g., BERT, RoBERTa) across established benchmark English and multilingual datasets for text classification and sequence labeling shows that encrypted models achieve performance parity with their original counterparts. This serves to safeguard performance, privacy, and security cohesively.
\end{abstract}

\section{Introduction}
In today's technology landscape, language models play a crucial role in numerous text-based applications. The rise of powerful pre-trained models like BERT \cite{devlin-etal-2019-bert}, RoBERTa \cite{liu2019roberta}, and GPT \cite{radford2018improving} and their successors, followed by a massive outburst of generative large language models such as GPT-3 \cite{brown2020language}, PaLM \cite{chowdhery2022palm}, LLaMa \cite{touvron2023llama} has led to a surge in innovative approaches and designs for large-scale language model pre-training. Characterized by an initial phase of broad pre-training, followed by targeted fine-tuning for specific contextual tasks, these models have found widespread utility across a diverse spectrum of applications \cite{otter2020survey,li2022pretrained} involving text classification, sequence labeling and text generation.

The widespread use of applications on server systems accessed through the internet has raised concerns about user privacy. Pre-trained language models are powerful tools, but their size and resource requirements make them challenging to use on personal devices. This leads to a situation where models are deployed on cloud-based servers and users send their text inputs over networks that can potentially be intercepted. For instance, sending text from users to servers using methods like HTTP GET/POST can expose the data to attacks known as \textit{Man-in-the-Middle (MitM)} attacks \cite{callegati2009man}, where an attacker sneaks between the user and the server, gaining access to and potentially altering the data being sent. This can result in data theft and misuse by malicious individuals. Another common worry related to using fine-tuned models on servers is the risk of unauthorized logging and misuse of users' personal data the organizations deploying these models \cite{o2003some,jones2005informed}. 

To deal with these problems, strong protective measures like encrypting data and models can be undertaken, involving potent one-way string encryption techniques such as the Secure Hash Algorithm (SHA) and passkey-based secure hash algorithms like \emph{Blake} \cite{fernandes2015implementation}. These methods convert input data into fixed-length hash values, which are incredibly difficult to reverse-engineer into the original input. However, encrypting input text introduces challenges for language models, typically trained and fine-tuned on plain text. This is due to two key reasons. Firstly, models pre-trained and fine-tuned on plain text may not adequately recognize encrypted inputs and may interpret them as \emph{out-of-vocabulary} (OOV) terms. Secondly, encryption hash algorithms provide enhanced security mostly by transforming text in a way that sacrifices a significant amount of linguistic information. For instance, consider the words \emph{cat} and \emph{cats} in plain text. They are closely related, sharing morphology and lexical semantics. However, their respective encrypted forms using a 256-bit SHA encryption, \texttt{77af778b} and \texttt{d936608b} (truncated for brevity), bear no clear relationship. This poses difficulties in suitably modeling language tasks with encrypted inputs. Pre-trained language models thrive on linguistic patterns in the input text, a task made challenging by the transformation introduced during encryption.

Our work focuses on making transformer-based language models compatible with encrypted inputs. Recognizing the impracticality of extensively pre-training language models with encrypted raw internet data, we choose a more efficient route. We first adapt a pre-trained language model (such as BERT) to comprehend key-based encrypted text and only fine-tune it on task-specific encrypted training data. During adaptation, we encode vocabulary entries using a secure key-based string encryption method, such as \emph{Blake} \cite{blake2}. We then shuffle the tokenizer's vocabulary and align the old token embeddings with the new vocabulary. This ensures that the text cannot be recovered either through network intrusion or server-side logging. Finally, to prevent retrieval of user inputs by reverse engineering of embeddings, we use distance-preserving geometric transformations of embeddings, which permanently alter embeddings to thwart attempts at reverting to original values.

After saving the altered model, we proceed with fine-tuning, where we encrypt the text examples in the training data using the same passkey used for the tokenizer transformation. The fine-tuning process follows usual established practices outlined in existing literature. Once fine-tuned and deployed, during inference, we propose that user text inputs first undergo pre-tokenization and encryption of text on users' devices, which is computationally lightweight. The encrypted text then reaches the server-based model pipeline, where a second-stage tokenizer further tokenizes the inputs for the adapted language models' inference.

We implement these adaptation and fine-tuning techniques on widely used pre-trained transformer based language models like BERT and RoBERTa, and evaluate these models on benchmark datasets for text classification and sequence labeling tasks. Our experiments demonstrate that the adapted and fine-tuned models achieve performance parity with the original models on relevant task metrics while ensuring enhanced privacy and security for the users.

Overall, this work focuses on a threat model identifying security and privacy risks, specifically interception vulnerabilities (unauthorized interception of user inputs) and privacy concerns (potential mishandling of user data). We prioritize irreversible data protection in fortification and fine-tuning, avoiding costly pre-training of the base model. The contributions can be summarized as follows: 

\begin{itemize}
    \item \textbf{Innovative Adaptation of Language Models:} We introduce a novel lightweight approach to easily adapt and fine-tune transformer based language encoders, enabling them to process passkey-encrypted inputs without requiring additional pre-training. This approach diverges from conventional practices in cybersecure language model development.
    \item{\textbf{Enhanced Text Security:}} We enhance text security in the adapted models by implementing systematic techniques such as tokenizer encryption, token shuffling, and token embedding transformation. These measures collectively make it challenging to reverse engineer the original text from the tokens and embeddings.
    \item{\textbf{Empirical Validation:}} Through extensive evaluation on various NLP benchmark datasets, we demonstrate that the overall transformation process maintains a lossless performance while being cost-effective. Additionally, the process is designed for easy replication to address new encryption requirements.
\end{itemize}

The code for this paper, is available at \texttt{\url{https://github.com/abhijitmishra/sentinellm-aaai2024}}.

\section{Related Work}
\label{sec:related}
Deep neural network language models like BERT, RoBERTa, and GPT have gained significant prominence in modern natural language processing. Yet, the layers within these architectures, particularly those involved in input processing such as embedding layers, introduce a potential risk of information leakage. In certain instances, attackers can exploit mathematical analysis based reverse engineering to recover 50-70\% of the original input \cite{song2020information}. Some recent studies have addressed such privacy vulnerabilities, but few efficiently tackle this issue without compromising model performance. \citeauthor{lee2022privacy}~(\citeyear{lee2022privacy}) propose a homomorphic encryption based method for privacy preserved text classification. \citeauthor{raeini2023privacy}~(\citeyear{raeini2023privacy}) explore data encryption based on polynomial spaces and matrices for privacy preserving language model access, where as \citeauthor{yu2021differentially}~(\citeyear{yu2021differentially}) propose differential privacy based solutions for the same. However, these approaches are not entirely immune to the risk of data leakage. Also for differential privacy based solutions, susceptibility arises from the fact that the definition of private data varies by company, and even data labeled as "non-private" by the company remains vulnerable to leakage. Furthermore, these methods are prone to potential performance losses. Our approach focuses on fully encrypting the data at the user's end, with little to no performance loss in most cases.

\begin{table*}[t]
    \centering
    \footnotesize
    \begin{tabular}{p{3.5cm} p{12.5cm}}
        \toprule
        \midrule
        \textbf{Raw Text} & List down the medicare benefits that are associated with Social Security Numbe 000-00-0000 and Dates of Birth 1st January, 1950. \\
        \textbf{Tokenized Text} & list down the medicare benefits that are associated with social security numb \#\#e 000 - 00 - 000 \#\#0 and dates of birth 1st january , 1950 \\
        \textbf{Encrypted Version1} & 53bb 9b57 a6b7 5325 9919 c438 0187 eec5 98e2 f3c6 cd0b 4ae4 6c6d 9a31 fb0e ae90 fb0e 9a31 a771 3ecd e353 5787 af69 5293 af69 5293 713f 0ee1 \\
        \textbf{Encrypted Version2} & 670c ba35 e816 eef6 382c 0bf7 baab 7563 0595 0c5a 1b5d e924 2c06 f9d8 8a5f 3f9b 8a5f f9d8 5faa 9881 7fc2 92c6 fda9 d228 fda9 d228 0cd1 25b2 \\
        \bottomrule
    \end{tabular}
    \caption{Examples showing input, tokenized and encrypted text. We show word-pieces from BertTokenizer. Version1 and Version2 are different encrypted hashes obtained using \textit{Blake-32} bit encryption with example passkeys \texttt{llm123} and \texttt{nlp2023} respectively. Hashes are truncated for brevity.}
    \label{tab:example_blake_table}
\end{table*}
Training smaller, private models based on large pre-trained models has become another prominent approach in privacy-preserving model development. Knowledge distillation has proven to be highly effective in this context. In their work, \citeauthor{wang2019private}~(\citeyear{wang2019private}) and \citeauthor{shejwalkar2021membership}~(\citeyear{shejwalkar2021membership}) adopt the knowledge distillation technique to safeguard member privacy in high-volume machine learning. Their focus, however, is on preserving the model's classification ability, thus making the solutions non-generic, unlike ours. Our work shares similarities with \citeauthor{qu2021natural}~(\citeyear{qu2021natural}), who introduce an embedding transformation technique involving noise induction and token character perturbation-based encryption. However, they apply it to a limited task set, and also need to continue pre-training  language models like BERT with a masked language modeling objective. In contrast, we perform post-hoc transformation of pre-trained language models, thus bypassing the resource-intensive process of extensive model pre-training. This makes our approach a first of its kind to the best of our knowledge.
\section{Central Idea}
\label{ref:central_idea}
Making sure that network-based transactions are secure and private involves a critical requirement: \emph{at no point during or after the transaction should the text provided by the user be understandable by people or machines to reveal any information about the text}. In usual situations, there are two ways security could be compromised: (a) a person sneaking into the network could intercept and collect personal data from the network packets, or (b) data reaching the server and being used by models for predictions might be stored and accessed for longer than necessary. Although these actions cannot be completely prevented, one can avoid security breaches by employing special encryption methods with \emph{one-way encryption} algorithms. These algorithms are intentionally designed to be slow, rendering it impossible to revert the encrypted data to its original form; in other words, they cannot be decrypted. The encrypted data, often called ciphertext, is a scrambled or hashed uninterpretable version of the input text. Well-known algorithms that are considered strong for encrypting sensitive data (like passwords) include the IBM's Data Encryption Standard (DES), the Secure Hash Algorithm family (SHA) \cite{penard2008secure}, MD5, and more recent methods like Blake \cite{blake2}. Algorithms like these typically help encrypt any length text into 128 or 256 bit hashes. 

While these encryption methods guarantee robust conversion of text into ciphertext, they often obscure the underlying linguistic signals when the entire input is encrypted. To find a middle ground, we propose encrypting the text at token levels, encrypting one token at a time. This allows language models to still learn from repeated tokens and patterns in training examples. This strategy, however, harbors a potential drawback. Encryption applied solely at the token level may fall short in providing substantial security, as attackers could still recover the original text by exploiting repeated usage, data logging, and pattern extraction via rudimentary means like regular expressions.

To tackle this, we propose utilizing keyed encryption methods, such as \textit{Blake}. This technique generates distinct ciphertext outputs based on user-specified passkeys or passphrases.  Illustrated in Table \ref{tab:example_blake_table}, you can see an example sentence and its tokenized version. The encryption process transforms the same input into two completely distinct encrypted results using two different passkeys. In practical scenarios, integrating keyed encrypted tokens in widely used language model tokenizers like WordPiece, SentencePiece \cite{kudo2018sentencepiece}, and BytePair Encoding \cite{shibata1999byte} is easy once a token-to-cipher map is established. This allows seamless use of pre-trained models on encrypted text, making them readily applicable. Yet, this doesn't guarantee safeguarding against server-side logging and reconstitution of text from cipher versions just based on token order. For instance, if the token indices in the original and encrypted vocabularies remain unchanged, recovering the original tokens becomes straightforward. This vulnerability can be mitigated by introducing random shuffling of tokens, disrupting their original positions in the vocabulary and other components of the language model reliant on token indices, such as token embedding layers of transformer-based models.
\begin{figure*}[t]
    \centering
    \includegraphics[width=14cm, height=12.5cm]{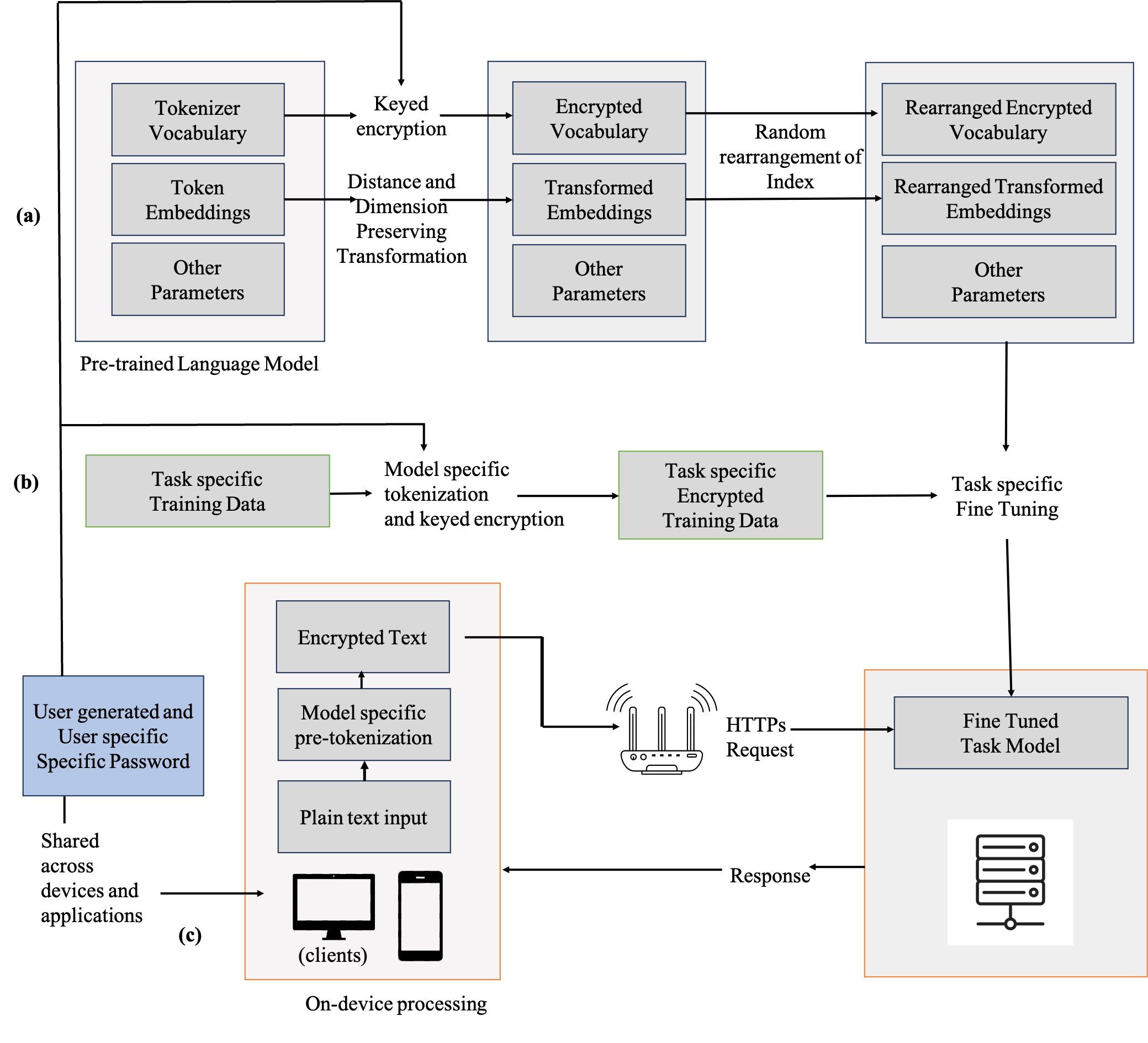}
    \caption{Illustration of the Workflow: User-Initiated Password-Driven Language Adaptation and Fine-Tuning Process. (a) Initial phase where a user-generated passkey initiates a one-time language adaptation. (b) Subsequent stage involving a one-time fine-tuning process. (c) Run-time scenario showcasing server-side inference on encrypted user input.}
    \label{fig:flowdiag}
\end{figure*}

Following token shuffling, the last layer of defense that we propose is towards ensuring that the retrieval of original tokens can't occur through accessing and reverse-engineering \emph{token embeddings}. Token embeddings serve as one of the earliest and potentially final layers of information accessible on the server side, directly conveying token-related details. In an ideal setting, token embeddings, renowned for capturing the distributional relationships between words \cite{levy-etal-2015-improving,mikolov2013efficient}, can inadvertently expose word information when subjected to certain mathematical operations like vector similarity based ranking and retrieval techniques. To address this concern, we propose a straightforward spatial and geometric transformation of token embeddings. This transformation preserves  crucial semantic fidelity and token semantics in the altered space, while strategically ensuring that the original embedding space remains indecipherable from the transformed one. We believe, this approach will help prevent the reversal of the original tokens from the transformed embeddings, bolstering the overall security measures.

We bring these ideas together in the upcoming section, where we explain how we adapt and refine pre-trained models using encrypted text.

\section{Model Adaptation and Fine Tuning}
\label{sec:lmadaptation}
Our approach is designed with personalized user security in focus, ensuring that data preparation and encryption are tailored for each individual user (or user groups). The entire process, encompassing pre-trained model adaptation, model fine-tuning, and runtime usage, is depicted in Figure \ref{fig:flowdiag}. We will now delve into the three key components of our methodology. It is important to note that in our discussion, we focus on a typical task setup: fine-tuning a pre-trained language model like BERT for a language-related task, such as text classification. This fine-tuning process involves utilizing an existing dataset for both training and evaluation purposes.
\subsection{Adaptation of Pre-trained Models}
\label{sec:adaptdetails}
This is a one time process and can be prompted by either a password update within an application or initiated by the server through periodic password changes. The process of adaptation entails the following three subprocesses.
\subsubsection{Tokenizer Encryption}
Initially, the tokenizer of the pre-trained model is modified by substituting its vocabulary items with their corresponding keyed encrypted forms using \textit{Blake} that generates a 32 bit encrypted output. This approach enables the creation of multiple tokenizers, each tailored for various users and groups, and these tokenizers can be stored on servers. Tokenizers are resource-efficient to store and manage, making it cost-effective to maintain multiple versions of the same foundational tokenizer, each with distinct encryption characteristics. Since model-specific tokenizers operate on-device and at server side, we implement a generic tokenizer that processes the encrypted version of the pre-tokenized text, and this approach can work on a large range of vocabulary based tokenizers like word-piece and sentence-piece.
\subsubsection{Token Embedding Transformation}
After tokenizer encryption, the token embeddings of the pre-trained transformer-based language models are extracted and transformed into an entirely different but distance preserving space. Considering a token embedding matrix of $M \times E$ where $M$ is the number of tokens in the vocabulary and $E$ is dimension of the embeddings, we apply a repeated number of \emph{glide-reflections} on the matrix. 
\begin{figure}[t]
    \centering
    \includegraphics[width=8.4cm]{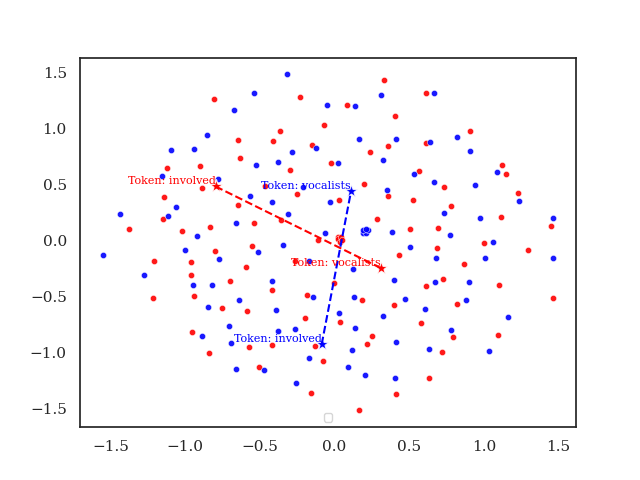}
    \caption{2D plot of embeddings for 100 random tokens from the original bert-base-uncased model (in red dots) and same tokens from the transformed model after 10 iterations of glide reflection (in blue dots). The plot further illustrates that sample tokens ``vocalists'' and ``involved'' have altered positions with  preserved spatial distance (dashed lines)}.
    \label{fig:glide}
\end{figure}
A glide reflection is a type of transformation that combines a reflection and a translation in Euclidean geometry. Given any line vector $l$ and a translation vector $t$ that have equal number of dimensions as an embedding vector $e$ (\textit{i.e.} $e,l,t  \in \mathbb{R}^d$ ), the reflection of the embedding vector $e$ can be given as:

\begin{equation}
\mathbf{e}_{\text{reflected}} = \mathbf{e} - 2 \cdot \frac{\mathbf{e} \cdot \mathbf{l}}{\mathbf{l} \cdot \mathbf{l}} \cdot \mathbf{l} \\
\label{eq:reflection}
\end{equation}

and following a translation, the glided transformation would result in 
\begin{equation}
\mathbf{e}_{\text{glide}} = \mathbf{e}_{\text{reflected}} + t\\
\label{eq:glide}
\end{equation}

Glide reflection holds significance in embedding transformations due to its ability to maintain the spatial distances between any pair of embedding vectors, labeled as $e_1$ and $e_2$, in Minkowski space. This preservation remains consistent regardless of the specific choices regarding the line of reflection or translation vector used. This is because both reflections and translations are operations that inherently keep distances unchanged.

While we acknowledge that a straightforward process like glide reflection does not provide an absolute assurance against the potential recovery of original vectors from the transformed space, it is reasonable to assume that employing a matrix of vectors undergoing a sequence of glide reflections around various random lines and translation vectors would greatly amplify the complexity of any attempt to recover the initial vectors. This intricate sequence of transformations makes the task of returning to the original vectors highly improbable. To illustrate glide reflection, take a look at Figure \ref{fig:glide}. This figure displays embeddings from the original BERT (\texttt{bert-base-uncased}) model for 100 random tokens in red, while the transformed token embeddings resulting from 10 glide reflection iterations (\textit{i.e.,} $nglide = 10$) are shown in blue. To fit the 768-sized embeddings into a 2D space for visualization, we used Multi Dimensional Scaling \cite{kruskal1978multidimensional}. As observed, even though two randomly chosen tokens, ``involved'' and ``vocalists'' have shifted from their original positions, the distances between them, measured in a straight line, have been maintained. This insight highlights the intriguing property of glide reflection. If this approach is to be adopted by multiple clients, the computational concern of allocating different passkeys and gliding values to each client is minimal as they involve deterministic transformations.
\begin{algorithm}[t]
\caption{AdaptLM}
\label{alg:encrypt}
\begin{algorithmic}[1]
\footnotesize
\Function{AdaptLM}{$M_{E}$, $nglides$, $V$, $key$}
  \State $d \gets $ dimension of E
  \For{$i \gets 1$ to $nglide$}
    \State $\text{l} \gets$ \Call{GenRandomVector}{$d$}
    \State $\text{t} \gets$ \Call{GenRandomVector}{$d$}
    \State $M_{E} \gets$ \Call{GlideReflection}{$M_{E}$, $l$, $t$}
  \EndFor
    \State $V \gets$ \Call{Blake2bEncryption}{$V$, $key$} 
    \State $r \gets$ shuffled indices from 0 to $d-1$ except special tokens 
    \State $V, M_{E} \gets$ \Call{Shuffle}{$V$, $M_{E}$,$r$}
  \State \Return $V$, $M_{E}$
\EndFunction

\Function{GlideReflection}{$M_{E}$, $\text{l}$, $\text{t}$}
    \State Initialize $M_{\text{reflected}} \gets$ to empty. 
    \For{vector $e \gets  M_E$}
        \State $e_{\text{reflected}} = e - 2 \cdot \frac{e \cdot e}{l \cdot e} \cdot l$
        \State Add $e_{\text{reflected}}$ to $M_{\text{reflected}}$
    \EndFor
    \State $M_E \gets M_{\text{reflected}} + t$
    \State \Return $M_E$
\EndFunction

\Function{Shuffle}{$V$, $M_{E}$, $r$}
  \State $V, M_{E} \gets$ Shuffled row indices of $M_{E}$ and vocabulary $V$ and based on $r$ 
  
  \State \Return $V, M_{E}$
\EndFunction
\end{algorithmic}
\end{algorithm}
\subsubsection{Vocabulary and Embedding Rearrangement}
After tokenization, encryption, and embedding transformations, the next step involves reorganizing the vocabulary and corresponding embedding indices while preserving alignment. This process is achieved by randomly selecting a subset of indices for shuffling, excluding specific tokens reserved for padding, classification, beginning and end of inputs. The shuffling and reassignment of indices occur within both the vocabulary list and the embedding matrix. 

After the reordering of indices, the adapted pre-trained model is stored and becomes available for future fine-tuning processes. The complete process is outlined in Algorithm \ref{alg:encrypt}. The inputs to the algorithm are the embedding matrix $M_E$, indexed vocabulary list $V$, number of iterations for glide reflection $nglide$ and the algorithm returns a transformed and shuffled version of $M_E$ and encrypted and shuffled vocabulary list $V$. 
\begin{table*}[t]
    \centering
    \footnotesize
    \begin{tabular}{lcccccccccc}
        \toprule
        \textbf{Model} & \textbf{CoLA} & \textbf{SST-2} & \textbf{MRPC} & \textbf{STS-B} & \textbf{QQP} & \textbf{MNLI} & \textbf{QNLI} & \textbf{RTE} & \textbf{WNLI} \\
        \midrule
        $\mathrm{BERT_{original}}$ & 56.53 & 92.32 & 88.9 & 88.6 & 90.7 & 83.9 & 90.7 & 65.7 & 56.3 \\
        $\mathrm{RoBERTa_{original}}$ & 67.8 & 94.8 & 90.2 & 91.2 & 91.9 & 87.6 & 92.8 & 78.7 & 56.34 \\
        $\mathrm{EBERT_{passkey=llm123, nglide=3}}$  & 55.3 & 92.55 & 89.76 & 88.61 & 91.14 & 84.43 & 91.45 & 66.06 & 56.34 \\
        $\mathrm{ERoBERTa_{passkey=llm123, nglide=10}}$  & 59.05 & 94.61 & 92.04 & 89.96 & 91.19 & 87.75 & 92.48 & 70.04 & 56.34 \\
        $\mathrm{EBERT_{passkey=nlp2023, nglide3}}$  & 55.24 & 92.66 & 89.64 & 88.69 & 91.19 & 84.56 & 91.43 & 66.79 & 56.34 \\
        $\mathrm{ERoBERTa_{passkey=nlp2023, nglide10}}$ & 60.57 & 94.38 & 92.14 & 89.74 & 91.40 & 87.62 & 92.71 & 68.95 & 56.34 \\
        \bottomrule
    \end{tabular}
    \caption{GLUE benchmark task metrics for the encryption adapted models starting with the letter $E$ with different passkeys and number of glides (nglides), specified in the subscripts. Metrics: CoLA - Matthews Correlation, SST-2 - Accuracy, MRPC - F1, STS-b - Pearson, QQP - Accuracy, MNLI - Matched Accuracy, QNLI - Accuracy, RTE - Accuracy, WNLI - Accuracy.}
    \label{tab:glueresults}
\end{table*}
\subsection{Fine-tuning}
With the model now prepared to accept and process encrypted inputs, we proceed to fine-tune it using encrypted task-specific training data. In the case of a pre-trained model and a tokenizer customized with a provided passkey, the fine-tuning process begins by tokenizing the training data with the original tokenizer (prior to encryption). These tokens are then isolated and encrypted using the \textit{Blake} keyed encryption algorithm. Using the encrypted tokens and existing labels from the dataset, models are fine-tuned for various classification and sequence labeling tasks.  In scenarios involving sequence labeling tasks, the integrity of the input and token mapping is preserved during the preprocessing of training data. It is worth noting that, given the similar configurations of the adapted and original models, fine-tuning demands equivalent time and resources as original models.
\subsection{Real-time Usage}
To accommodate the need for token-level encryption during inference, we present a straightforward approach as follows. First, the user's input is tokenized using a model-specific tokenizer. For instance, if the adapted model is derived from a bert-base model, the applicable tokenizer is BertTokenizer, which should execute on the user's device. Next, the tokenized input is encrypted using the user specific passkey. It is reasonable to anticipate that on-device tokenization and encryption are feasible, as these operations generally involve lightweight processes that do not require substantial computational resources. With on-device tokenization and encryption done, the encrypted data is transmitted through the network, adhering to security protocols, such as HTTPS.

Upon arriving at the server, the tokenizer corresponding to the adapted model takes over. This tokenizer further tokenizes the encrypted input and correlates the tokens with their respective indices. Given that indices and embeddings have been shuffled at this stage, the likelihood of order-dependent retrieval diminishes significantly.
\begin{table*}[t]
    \centering
    \footnotesize
    \begin{tabular}{lcccc}
        \toprule
        \textbf{Model} & Precision & Recall & F1 & Accuracy \\
        \midrule
        $\mathrm{BERT_{original}}$ & 94.69 & 95.15 & 94.92 & 98.93  \\
        $\mathrm{RoBERTa_{original}}$ & 95.34 & 95.98 & 95.66 & 99.25 \\
        $\mathrm{EBERT_{passkey=llm123, nglide=3}}$  & 93.86 & 94.87 & 94.36 & 98.90 \\
        $\mathrm{ERoBERTa_{passkey=llm123, nglide=10}}$  & 94.95 & 95.79 & 95.37 & 99.23 \\
        $\mathrm{EBERT_{passkey=nlp2023, nglide3}}$  & 93.86 & 94.68 & 94.27 & 98.85 \\
        $\mathrm{ERoBERTa_{passkey=nlp2023, nglide10}}$ & 95.16 & 95.96 & 95.56 & 99.24 \\
        \bottomrule
    \end{tabular}
    \caption{Classification Metrics of Models on Sequence Labeling Task (NER) on CoNLL2003 Dataset}
    \label{tab:nerresults}
\end{table*}
\begin{table*}[t]
    \centering
    \footnotesize
    \begin{tabular}{lccc}
        \toprule
        \textbf{In-language} & es-es & de-de & zh-zh \\
        \midrule
        $\mathrm{BERTMultilingual_{original}}$ & 77.3 & 75.2 & 74.2 \\
        $\mathrm{EBERTMultilingual_{passkey=llm123, nglide=10}}$ & 78.7 & 76.8 & 73.9 \\
        \midrule
        \textbf{Zero-shot} & en-es & en-de & en-zh \\
        \midrule
        $\mathrm{BERTMultilingual_{original}}$ & 74.3 & 70.5 & 63.8 \\
        $\mathrm{EBERTMultilingual_{passkey=llm123, nglide=10}}$ & 75.5 & 71.1 & 65.3 \\
        \bottomrule
    \end{tabular}
    \caption{Accuracy of Models Multilingual Natural Language Inference (XNLI) Task}
    \label{tab:xnliresults}
\end{table*}
Following tokenization and embedding lookup, the input undergoes the model's deeper layers in the fine-tuned architecture and the resulting decisions are relayed back to the user in an appropriate response structure. This streamlined approach ensures data security while maintaining efficiency, offering a robust solution for token-level encryption during inference.
\section{Experimental Setup}
\label{sec:expsetup}

\subsection{Datasets and Language models}
With the objective of demonstrating equivalence between fine-tuned original models and fine-tuned adapted models (as detailed in Section \ref{sec:lmadaptation}), we opt for established and extensively documented datasets for text classification. These datasets include the well-regarded General Language Understanding Evaluation (GLUE) benchmark dataset \cite{wang2018glue}, the CoNLL2003 Named Entity Recognition Dataset \cite{sang2003introduction} for sequence labeling tasks, and the XNLI dataset \cite{conneau2018xnli} specifically curated for multilingual natural language inference evaluation. For XNLI, we evaluate all models in two settings: (a) \textit{In-language}, where the data for training and testing come from the same language source and (b) \textit{Zero-shot}, where the training data contains only English examples and test data is in target language. 
\subsection{Implementation Details}
For the pre-trained models, we select BERT (\texttt{bert -base-uncased}), RoBERTa (\texttt{roberta-base}), and mBERT (\texttt{bert-base-multilingual-uncased}). We source these pretrained models and their corresponding fine-tuning configurations from Huggingface\footnote{\url{https://huggingface.co}}, a well-known platform in the field. For adaptation, we consider two sample passkeys: ``llm123'' and ``nlp2023''. To ensure reproducibility, all randomized operations are initiated using numbers obtained from the passkeys through character-to-byte conversion. The line and translation vectors are obtained by randomly initializing vectors with non-zero positive values between $0$ and $1$. We explore the effects of different numbers of gliding iterations: $3$ and $10$. For keyed encryption, \textbf{Blake2} implementation of the \texttt{hashlib} Python library is used. 

\section{Results and Analysis}
\label{sec:results}
Table \ref{tab:glueresults} showcases the results obtained across various tasks within the GLUE benchmark. Our diverse set of adapted and fine-tuned model variants consistently demonstrates performance within a 2\% margin of the original models\footnote{Result references: \url{https://gluebenchmark.com/leaderboard/}, \url{https://github.com/facebookresearch/fairseq/tree/main/examples/roberta}, \url{https://huggingface.co/textattack/roberta-base-WNLI}} across tasks, with the exception of CoLA and RTE (with RoBERTa). A specific observation is the performance parity achieved for the WNLI task by specifying a slightly different learning rate (from the default value of 2e-5 to 1e-4). Similar parity is evident in the NER task for both BERT and RoBERTa models. Additionally, for the multilingual XNLI task, our adapted fine-tuned models demonstrate accuracy parity when compared to the original models across both in-language and zero-shot settings. Notably, models subjected to a higher number of transformation iterations ($nglide=10$) exhibit similar performance as the original model while ensuring better privacy preserving abilities due to repeated embedding transformations, as discussed in Section \ref{sec:adaptdetails}. 

In instances where models display lower performance, we speculate that refining the remaining layers of the language model may require extending the training iterations--a hypothesis currently under validation. This highlights an ongoing effort to optimize model performance and underscores the importance of iterative refinement in achieving robust language understanding across diverse tasks within the GLUE benchmark. It is also worth noting that difference in results with identical models with different pass keys arise from using distinct random seeds during adaptation. Consistent outcomes can be achieved by employing the same seed values.

\subsection{Investigations on Recoverability of Tokens}
We investigate the feasibility of retrieving tokens by leveraging embeddings from both the original and transformed versions. To explore this, we calculate the pairwise Euclidean distances between original and transformed token embeddings, subsequently compiling a list of closest neighbors for each vocabulary token. We then determine the accuracy of predicting the nearest neighbor (with neighbors obtained from the original embeddings considered as the benchmark). The accuracies turn out to be negligible (\textbf{0.009\%} in most cases), illustrating a systematic shift of embeddings from their original positions.

Given that distance preservation occurs in the transformations applied to token embeddings during adaptation, there also arises a valid concern that pairwise distances might serve as clues for recovering embeddings. While this holds true for non-fine-tuned models, our primary focus is to ensure the infeasibility of such recovery once the models undergo fine-tuning, as these fine-tuned models are the ones ultimately deployed for practical use. To validate this, we execute a comprehensive experiment. Both the original and adapted models for BERT and RoBERTa are fine-tuned using only $10$ examples and only for one epoch across all GLUE tasks. After fine-tuning, we extracted the embedding matrices and generated ranked lists of tokens based on pairwise distances for each token in the vocabulary. Remarkably, the rankings obtained for the original fine-tuned models and their adapted counterparts exhibit only about a \textbf{0.01\%} match across all tasks. This substantial disparity underscores the inefficacy of potential attacks aimed at recovering tokens. Overall, these experiments serve as a crucial assurance of the security and integrity of our fine-tuned models in real-world deployment scenarios.
\section{Conclusion and Future Work}
\label{label:conclusion}
In this paper, we addressed pressing privacy and security concerns associated with pre-trained transformer-based language models, emphasizing the substantial risks posed by server-deployed models for user privacy. To mitigate these challenges, we proposed an innovative approach involving the adaptation and fine-tuning of existing language models for passkey-encrypted user-specific text. Adaptation of pre-trained language models through irreversible transformations on the tokenizer and token embeddings facilitates secure inference on encrypted inputs, ensuring the confidentiality of text. Furthermore, we observed that fine-tuning on encrypted training data maintains performance parity with original models across various benchmark datasets. In the future, we aim to expand our approach to include generative models and investigate improved encryption strategies for non-English and multilingual tasks. As encryption requirements evolve, our method can seamlessly incorporate advanced cryptographic algorithms or explore post-quantum encryption methods. While we consider our approach robust against common attacks, additional safeguards like multi-party computation or homomorphic encryption could strengthen its security, marking a substantial advancement in ensuring user privacy and security in practical applications.

\section*{Ethical Statement}
\label{sec:ethics}
Although our primary objective was to establish secure adaptive solutions for any transformer-based language model, it is essential to note that our approach has been evaluated exclusively on fine-tuned language encoders and not on generative language models. Generative models heavily capitalize on shared tokenizers and vocabulary for both encoders and decoders, which is incompatible with our approach. Also, while we are advancing privacy-preserving methodologies, it is crucial to emphasize that our endeavors have not involved the utilization of private data at any point. Our approach has exclusively relied on publicly accessible benchmark datasets, notably including GLUE. However, it's essential to acknowledge the potential biases present in these public datasets, particularly within the realm of language model benchmarking.
\bibliography{aaai24}

\end{document}